%% file: main.tex
\documentclass[10pt,twocolumn]{article}

\usepackage[top=0.75in,bottom=0.85in,left=0.7in,right=0.7in]{geometry}
\usepackage{amsmath,amssymb,amsfonts}
\usepackage{booktabs}
\usepackage{graphicx}
\usepackage{microtype}
\usepackage{cite}
\usepackage{url}
\usepackage{xcolor}
\usepackage{listings}
\usepackage{caption}
\usepackage{flushend}
\usepackage{pgfplots}
\pgfplotsset{compat=1.18}
\usetikzlibrary{arrows.meta,positioning,fit,backgrounds,shadows,calc}
\usepackage[ruled,vlined,linesnumbered]{algorithm2e}
\usepackage[colorlinks=true,linkcolor=blue,citecolor=blue,urlcolor=blue]{hyperref}

\lstdefinelanguage{Rust}{
  keywords={pub, struct, fn, let, mut, match, if, else, while, for, in, return, break, loop, continue, impl, trait, type, const, static, use, mod, as, ref, where, true, false, Some, None, Ok, Err, Arc, Mutex, AtomicU64, Ordering},
  keywordstyle=\color{blue}\bfseries,
  ndkeywords={self, Self, i64, u64, u32, usize, String, str, bool},
  ndkeywordstyle=\color{teal}\bfseries,
  sensitive=true,
  comment=[l]{//},
  morecomment=[s]{/*}{*/},
  commentstyle=\color{gray}\ttfamily,
  stringstyle=\color{purple}\ttfamily,
  morestring=[b]",
}

\title{\Large\textbf{An Edge-First Multimodal Asset Management Studio and Retrieval Engine for Heterogeneous AI-Generated Media}}

\author{
  \textbf{Chaoyu Lin}\\[0.3em]
  \normalsize Independent Researcher\\[0.2em]
  \normalsize \href{mailto:berrywahlberg@gmail.com}{berrywahlberg@gmail.com}
}

\date{}

\begin{document}

\maketitle

\input{sections/00_abstract}

\input{sections/01_introduction}

\input{sections/02_related_work}

\input{sections/03_architecture}

\input{sections/04_key_mechanisms}

\input{sections/05_evaluation}

\input{sections/06_open_source}

\input{sections/07_conclusion}

\end{document}

%% file: sections/00_abstract.tex
\begin{abstract}
AI-generated image collections combine visual content with heterogeneous provenance such as prompts, model parameters, and workflow graphs. Managing these collections locally requires coordinated metadata extraction, responsive browsing, and retrieval without mandatory remote inference. We present \textbf{Berry AI Studio}, an edge-first asset management system built with Vue, Tauri, Rust, SQLite, and ONNX Runtime. Berry integrates format-specific metadata parsing with raw-payload retention, generation-aware cancellation of queued thumbnail work, a persistent multi-tier thumbnail manifest, cursor pagination, and local image--text retrieval. We evaluate these mechanisms in five independent repetitions on a pinned source revision. With 50,000 catalog records, the median warm first-page database call was 285.3~ms; at offset 40,000, keyset pagination reduced median raw-query latency from 57.9 to 0.46~ms. A controlled browser harness recorded a 0.3~ms median-of-run P95 update time during virtualized scrolling, while cancellation prevented all 1,200 stale speculative decodes in the tested trace. A versioned 10-case corpus produced 53/53 exact metadata-field matches, and controlled retrieval and grouping tests exercised the authentic storage and clustering paths. The implementation, benchmark scripts, raw logs, and checksummed result packages are publicly available.
\end{abstract}


%% file: sections/01_introduction.tex
\section{Introduction}
\label{sec:intro}

Latent diffusion models~\cite{rombach2022high} and generation interfaces such as ComfyUI~\cite{comfyui2023} and Stable Diffusion WebUI~\cite{sdwebui2022} support iterative image production through prompts, seeds, model choices, and conditioning parameters. The resulting files carry both visual content and generation provenance. Organizing these outputs therefore involves more than locating an image: users may also need to recover its prompt, identify a model dependency, or inspect the workflow that produced it.

\subsection{Statement of Need}
\label{sec:statement_of_need}
This work addresses three engineering requirements for local generative-media libraries.
\begin{enumerate}
    \item \textbf{Queryable provenance.} Generation tools serialize metadata differently. WebUI parameter strings and ComfyUI graph JSON require distinct parsers before prompts and model attributes can be exposed through a common index. Retaining source payloads is also useful when normalized fields omit tool-specific information.
    \item \textbf{Viewport-aware background work.} Virtualization limits the number of rendered elements, but does not by itself prevent queued thumbnail requests from outliving their visible targets. Rapid scrolling can leave native workers processing images that are no longer needed. Bounding concurrency and checking request relevance before decoding can reduce this work.
    \item \textbf{Local retrieval.} Creators handling unreleased assets may prefer to keep media and prompts on their own devices. Local embedding inference supports this requirement, while introducing resource and model-distribution constraints that must be considered alongside interface responsiveness.
\end{enumerate}
These requirements motivate Berry's design; they do not establish that all existing asset managers lack the corresponding capabilities. A controlled comparison with other products is outside the present evaluation.

\subsection{Contributions}
\label{sec:contributions}
Berry is a systems integration contribution rather than a new representation-learning method or database primitive. This report presents:
\begin{itemize}
    \item A modular local architecture that combines heterogeneous metadata extraction, indexed catalog storage, and on-device image--text retrieval.
    \item A thumbnail pipeline that associates requests with viewport generations, checks for stale work before decoding, and reuses cached resolution tiers through a persistent manifest.
    \item An implementation-grounded account of cursor pagination and prompt-based grouping, including their assumptions and limitations.
    \item A reproducible seven-part evaluation covering catalog ingestion, pagination, cache reconciliation, browser interaction, cancellation, metadata extraction, prompt grouping, and vector retrieval, with raw logs and checksummed artifacts for five independent repetitions.
\end{itemize}

%% file: sections/02_related_work.tex
\section{Related Work}
\label{sec:related_work}

\subsection{Generative Platforms and Provenance}
Stable Diffusion~\cite{rombach2022high} provides a foundation for generation tools with different interaction and serialization models. Stable Diffusion WebUI~\cite{sdwebui2022} exposes parameter-oriented workflows, whereas ComfyUI~\cite{comfyui2023} represents workflows as connected nodes. Fooocus~\cite{fooocus2023} offers another generation interface. Berry's metadata layer accommodates these differences through format-specific extraction and a shared representation. Preserving a source payload and extracting selected searchable fields are separate operations; neither implies complete semantic interpretation of arbitrary custom nodes.

\subsection{Digital Asset Management}
Digital asset managers such as digiKam~\cite{digikam2010} and Eagle~\cite{eagle2020} provide relevant context for organizing image collections. Berry focuses on generation provenance and its integration with local retrieval and thumbnail scheduling. We do not infer competitors' internal scheduling or storage algorithms from their user-facing features, and do not claim comparative performance without matched workloads and version-specific measurements.

\subsection{Virtualization and Background Scheduling}
List virtualization reduces presentation work by rendering a viewport and an overscan region. Native thumbnail processing requires a complementary policy because reducing the visible element count does not automatically cancel queued I/O. Berry combines a bounded Rayon pool~\cite{rayon2023} with generation checks at request and worker boundaries. This is cooperative cancellation: it can skip stale work before decoding begins, but does not interrupt a decoder that is already running.

\subsection{Local Multimodal Retrieval}
CLIP~\cite{radford2021learning} and SigLIP~\cite{zhai2023sigmoid} learn image--text representations that support cross-modal retrieval. Berry integrates these model families through ONNX Runtime~\cite{onnxruntime2021}. Its contribution lies in deploying retrieval within a local catalog workflow, rather than modifying the models' training objectives. Retrieval quality, indexing cost, and inference latency depend on the model and hardware; the microbenchmarks in this report do not evaluate those dimensions.

%% file: sections/03_architecture.tex
\section{System Architecture and Implementation}
\label{sec:architecture}

Berry separates interactive presentation from filesystem operations, database access, and model inference. Figure~\ref{fig:architecture} summarizes the three principal layers: a Vue/TypeScript frontend, a Tauri~2.0 native bridge~\cite{tauri2024}, and specialized Rust crates~\cite{matsakis2014rust}. This separation provides explicit boundaries for request dispatch, persistence, and background processing.

\input{figures/fig_architecture}

\subsection{Presentation and Virtualization}
Grid and masonry views select visible items with an overscan margin. For a masonry layout, cached geometry is partitioned by column, allowing a binary search for the first visible item followed by enumeration of intersecting items. For $C$ columns and $V$ returned items, this selection costs approximately $\mathcal{O}(C\log(N/C)+V)$ under balanced column occupancy; constructing the geometry remains a separate operation over the collection. Scroll updates are coalesced with \texttt{requestAnimationFrame}.

Thumbnail object URLs have a cache lifecycle that includes revocation on eviction. An entry limit bounds retained references, but does not by itself bound decoded image memory or prove the absence of leaks. Section~\ref{sec:evaluation} distinguishes the cache simulation from browser-level memory measurements.

\subsection{Native Request Coordination}
Tauri exposes Rust commands to the frontend through IPC. Thumbnail commands update a shared atomic generation tracker:
\noindent\begin{minipage}{\columnwidth}
\begin{lstlisting}[language=Rust,basicstyle=\ttfamily\scriptsize]
generation_tracker.fetch_max(generation, Ordering::AcqRel);
\end{lstlisting}
\end{minipage}

Workers compare their request generation with the current value before expensive decoding. Atomic coordination avoids acquiring a database mutex for this check; it does not remove locking elsewhere in the application.

\subsection{Backend Responsibilities}
The following five crates organize the mechanisms discussed in this report:
\begin{itemize}
    \item \textbf{\texttt{berry-domain}} defines shared entities and pure prompt-grouping logic.
    \item \textbf{\texttt{berry-metadata}} parses supported container metadata and generation formats into selected normalized fields while retaining raw payloads where supported.
    \item \textbf{\texttt{berry-storage}} implements SQLite persistence~\cite{hipp2020sqlite}, WAL configuration, query construction, and versioned migrations. The thumbnail manifest was introduced in migration~11; later migrations add cursor indexes and other capabilities.
    \item \textbf{\texttt{berry-scan}} coordinates directory scanning, change handling, and thumbnail processing in bounded Rayon pools~\cite{rayon2023}.
    \item \textbf{\texttt{berry-clip}} integrates ONNX Runtime~\cite{onnxruntime2021} for local image and text embeddings using CLIP/SigLIP model families~\cite{radford2021learning,zhai2023sigmoid}.
\end{itemize}
These module boundaries describe implementation responsibilities, rather than guarantees about whole-application memory usage or execution time.

%% file: figures/fig_architecture.tex
% TikZ System Architecture Diagram for Berry AI Studio
\begin{figure*}[t]
\centering
\resizebox{\textwidth}{!}{%
\begin{tikzpicture}[
    font=\sffamily\small,
    box/.style={draw=gray!45, line width=0.45pt, rounded corners=3pt, fill=white, inner sep=6pt},
    header/.style={font=\bfseries\sffamily\small, text=gray!80!black},
    subhead/.style={font=\sffamily\footnotesize, text=gray!70!black},
    crate/.style={draw=blue!55!black, line width=0.5pt, fill=white, rounded corners=2.5pt, inner sep=4.5pt, text width=2.55cm, minimum height=4.15cm, align=left},
    arrow/.style={-{Stealth[scale=0.85]}, line width=0.65pt, draw=blue!70!black},
    syncarrow/.style={-{Stealth[scale=0.85]}, line width=0.65pt, densely dashed, draw=red!70!black},
    lbl/.style={fill=white, inner sep=2pt, font=\tiny\sffamily, text=gray!90!black, rounded corners=1pt},
    redlbl/.style={fill=white, inner sep=2pt, font=\tiny\bfseries\sffamily, text=red!75!black, rounded corners=1pt}
]

% 1. Presentation Tier
\node[box, fill=blue!4, draw=blue!35, text width=16.0cm] (pres) {
    \begin{minipage}{15.8cm}
    \textbf{\textcolor{blue!80!black}{Presentation Tier (Vue 3 / TypeScript / Vite)}} \hfill \textit{\footnotesize Viewport Virtualization}\\[3pt]
    \begin{tabular}{@{}p{4.98cm} p{4.98cm} p{4.98cm}@{}}
    \textbf{Viewport Renderers} & \textbf{Layout \& Scroll Virtualization} & \textbf{Memory \& Concurrency} \\
    $\bullet$ Responsive \texttt{VirtualGrid} & $\bullet$ Column-indexed binary search & $\bullet$ Monotonic \texttt{generation} tracker \\
    $\bullet$ Masonry Waterfall Layout & $\bullet$ Animation-frame coalescing & $\bullet$ 3,000-item Blob URL LRU cache \\
    $\bullet$ Multi-column Table View & $\bullet$ 2-row bounded overscan margin & $\bullet$ Deterministic URL revocation \\
    \end{tabular}
    \end{minipage}
};

% 2. IPC Bridge
\node[box, fill=teal!4, draw=teal!45, text width=16.0cm, below=0.65cm of pres] (ipc) {
    \begin{minipage}{15.8cm}
    \textbf{\textcolor{teal!80!black}{Tauri 2.0 Native IPC Bridge (\texttt{src-tauri/src/commands.rs})}} \hfill \textit{\footnotesize High-Throughput Typed Serialization}\\[3pt]
    \begin{tabular}{@{}p{4.98cm} p{4.98cm} p{4.98cm}@{}}
    \textbf{Typed Command Dispatch} & \textbf{Lock-Free State Synchronization} & \textbf{Streaming Channels} \\
    $\bullet$ Serde JSON deserialization & $\bullet$ \texttt{thumbnail\_generation} & $\bullet$ Async directory scan progress \\
    $\bullet$ Native WebView & $\bullet$ \texttt{fetch\_max(gen, AcqRel)} & $\bullet$ Batch thumbnail event emitting \\
    \end{tabular}
    \end{minipage}
};

% 3. Crate Ecosystem Tier Anchor
\node[below=1.8cm of ipc] (crates_anchor) {};

% Crate 1: Domain (Far Left: x = -6.4cm)
\node[crate, at={(crates_anchor.center)}, xshift=-6.4cm, anchor=north] (c_domain) {
    \textbf{\texttt{\textcolor{blue!90!black}{berry-domain}}}\\[2pt]
    \scriptsize\textcolor{gray!80!black}{Zero-I/O Pure Core}
    \vspace{3pt}\hrule\vspace{3pt}
    \scriptsize
    $\bullet$ \texttt{ImageFile}, \texttt{Container}\\
    $\bullet$ \texttt{ExtractedMetadata}\\
    $\bullet$ \texttt{DetectedLora}\\
    $\bullet$ Set Jaccard:
    \[ J = \frac{|A\cap B|}{|A\cup B|} \]
    $\bullet$ Burst stack planning
};

% Crate 2: Metadata (x = -3.2cm)
\node[crate, at={(crates_anchor.center)}, xshift=-3.2cm, anchor=north] (c_meta) {
    \textbf{\texttt{\textcolor{blue!90!black}{berry-metadata}}}\\[2pt]
    \scriptsize\textcolor{gray!80!black}{Metadata Parsing}
    \vspace{3pt}\hrule\vspace{3pt}
    \scriptsize
    $\bullet$ Magic-byte sniffer\\
    $\bullet$ PNG \texttt{tEXt}/\texttt{iTXt} parsing\\
    $\bullet$ ComfyUI node parsing\\
    $\bullet$ A1111 parameter strings\\
    $\bullet$ LoRA tag/weight normalization
};

% Crate 3: Storage (x = 0cm)
\node[crate, at={(crates_anchor.center)}, anchor=north] (c_storage) {
    \textbf{\texttt{\textcolor{blue!90!black}{berry-storage}}}\\[2pt]
    \scriptsize\textcolor{gray!80!black}{Relational Persistence}
    \vspace{3pt}\hrule\vspace{3pt}
    \scriptsize
    $\bullet$ SQLite WAL mode\\
    $\bullet$ Versioned migrations\\
    $\bullet$ Cursor seek query:
    \[ \mathcal{O}(\log N) \text{ B-tree seek} \]
    $\bullet$ Manifest table\\
    $\bullet$ Change journal
};

% Crate 4: Scan (x = +3.2cm)
\node[crate, at={(crates_anchor.center)}, xshift=3.2cm, anchor=north] (c_scan) {
    \textbf{\texttt{\textcolor{blue!90!black}{berry-scan}}}\\[2pt]
    \scriptsize\textcolor{gray!80!black}{Parallel I/O \& Codecs}
    \vspace{3pt}\hrule\vspace{3pt}
    \scriptsize
    $\bullet$ Bounded Rayon pool\\
    $\bullet$ Pre-decode generation check\\
    $\bullet$ Image downsampling\\
    $\bullet$ WebP encoding\\
    $\bullet$ Manifest reconcile/evict
};

% Crate 5: CLIP (x = +6.4cm)
\node[crate, at={(crates_anchor.center)}, xshift=6.4cm, anchor=north] (c_clip) {
    \textbf{\texttt{\textcolor{blue!90!black}{berry-clip}}}\\[2pt]
    \scriptsize\textcolor{gray!80!black}{Edge ML Inference}
    \vspace{3pt}\hrule\vspace{3pt}
    \scriptsize
    $\bullet$ ONNX Runtime sessions\\
    $\bullet$ SigLIP/CLIP models\\
    $\bullet$ Local-only execution\\
    $\bullet$ 512/768-d normalization\\
    $\bullet$ Cosine ranking
};

% Native tier frame uses the exact horizontal bounds of the two tiers above.
\coordinate (native_nw) at ([yshift=0.62cm]c_domain.north -| pres.west);
\coordinate (native_se) at ([yshift=-0.18cm]c_domain.south -| pres.east);
\begin{scope}[on background layer]
  \path[draw=blue!35, line width=0.45pt, fill=blue!2, rounded corners=3pt]
    (native_nw) rectangle (native_se);
\end{scope}
\node[anchor=north west, font=\bfseries\sffamily\footnotesize, text=blue!85!black]
    at ([xshift=0.16cm,yshift=-0.12cm]native_nw)
    {Native Crates (\texttt{crates/*})};

% Connecting Arrows: Presentation Tier <-> IPC Bridge
\draw[arrow] ([xshift=-3.5cm]pres.south) -- ([xshift=-3.5cm]ipc.north)
    node[midway, lbl] {IPC Invocations};
\draw[arrow] ([xshift=3.5cm]ipc.north) -- ([xshift=3.5cm]pres.south)
    node[midway, lbl] {Events \& Results};

% Connecting Arrows: IPC Bridge -> Crates
\draw[arrow] ([xshift=-3.2cm]ipc.south) -- (c_meta.north)
    node[midway, lbl] {Parse Task};

\draw[arrow] ([xshift=0cm]ipc.south) -- (c_storage.north)
    node[midway, lbl] {Keyset Query};

% Disjoint arrows to c_scan
\draw[syncarrow] ([xshift=2.3cm]ipc.south) -- ([xshift=-0.6cm]c_scan.north)
    node[pos=0.35, left, redlbl] {Cancellation};

\draw[arrow] ([xshift=3.9cm]ipc.south) -- ([xshift=0.6cm]c_scan.north)
    node[pos=0.65, right, lbl] {Batch Decode};

\draw[arrow] ([xshift=6.4cm]ipc.south) -- (c_clip.north)
    node[midway, lbl] {Embeddings};

\end{tikzpicture}%
}
\caption{System architecture of Berry AI Studio. The architecture decouples the virtualized presentation tier from native operating system facilities via a typed Tauri 2.0 bridge and five modular Rust crates, coordinating cache lifecycles and cooperative cancellation.}
\label{fig:architecture}
\end{figure*}
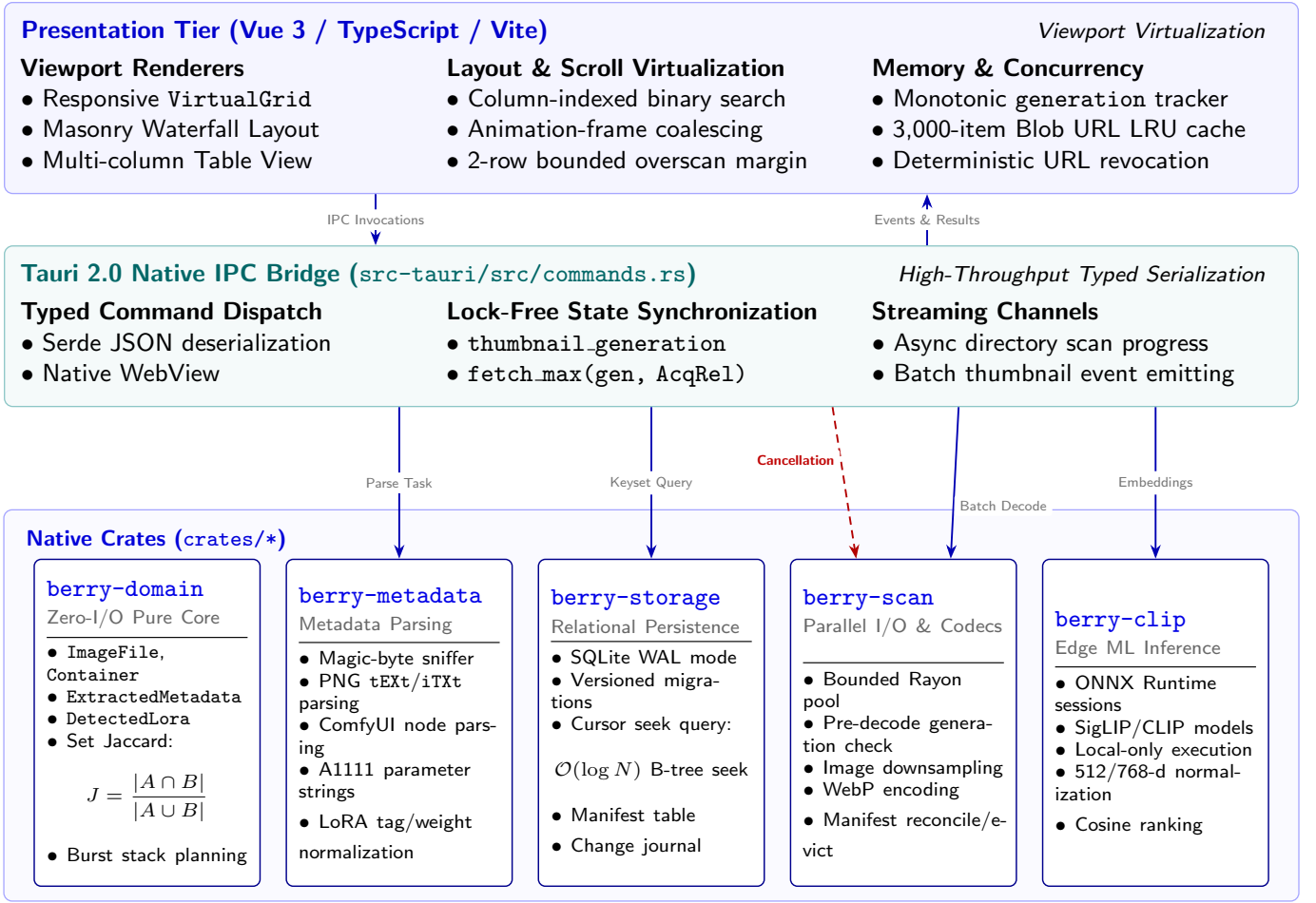

%% file: sections/04_key_mechanisms.tex
\section{Key Mechanisms and Algorithmic Design}
\label{sec:mechanisms}

This section details the primary algorithmic formulations and architectural mechanisms implemented in Berry AI Studio.

\subsection{Heterogeneous Metadata Extraction}
\label{sec:metadata_parsing}
Generation tools expose different metadata representations. Berry uses format-specific parsers to populate common fields such as prompts, seeds, sampler names, and model names. Raw-payload retention preserves information beyond these fields where supported; it is distinct from lossless semantic normalization.

\paragraph{ComfyUI graph handling.}
The parser accepts node-map JSON, wrapped prompt JSON, and workflow node arrays. For node maps, it identifies supported sampler classes and follows positive and negative links to text inputs. The text resolver handles direct text, dual text fields, and a supported upstream conditioning link. Model and dimension fields are also extracted from recognized node classes. This is partial interpretation of known structures, not general execution or reconstruction of an arbitrary workflow. Disconnected nodes and custom node semantics may make field attribution ambiguous. Cyclic malformed input requires separate robustness testing; a directed acyclic graph does not contain cycles by definition.

\paragraph{LoRA normalization.}
Separate extraction routines normalize supported inline LoRA references and graph metadata. Path and extension normalization help reconcile equivalent names. These operations do not establish that all model dependencies or their execution semantics can be recovered.

\subsection{Viewport-Generation-Aware Concurrency Scheduling}
\label{sec:viewport_concurrency}
When a user rapidly drags the scrollbar across tens of thousands of items, traditional image loading queues accumulate hundreds of off-screen requests. Decoding and downsampling require CPU time and temporary buffers, so obsolete requests can compete with currently visible work.

Berry reduces unnecessary queued work through cooperative, generation-aware cancellation implemented within bounded Rayon~\cite{rayon2023} thread pools:

\paragraph{Thread Pool Clamping:}
The dedicated thumbnail pool limits concurrent workers according to:
\begin{equation}
    \mathcal{T}_{\text{workers}} = \operatorname{clamp}\left(\left\lfloor \frac{N_{\text{logical\_cpus}}}{2} \right\rfloor, 2, 6\right)
\end{equation}
This bounds the dedicated Rayon pool (\texttt{THUMB\_POOL}) between 2 and 6 worker threads, limiting background parallelism without reserving or pinning physical cores. The lower bound may exceed the logical CPU count on very small systems.

\begin{algorithm}[t]
\caption{Generation-aware thumbnail request (simplified control flow)}
\label{alg:viewport_sched}
\KwIn{Shared generation $G$, request generation $g$, source $P$, file ID $f$, revision $m$, tier $\tau$, disk budget $B$}
\KwOut{Thumbnail path or cancellation/error}
$G \leftarrow \operatorname{atomicFetchMax}(G,g)$\;
\If{$\operatorname{loadAcquire}(G) \ne g$}{\Return cancellation\;}
$p \leftarrow \operatorname{ExactCachePath}(f,m,\tau)$\;
\If{$\operatorname{Exists}(p)$}{
    $\operatorname{RecordAccessIfDue}(p)$\;
    \Return $p$\;
}
$p \leftarrow \operatorname{FindExistingSufficientTier}(f,m,\tau)$\;
\If{$p \ne \emptyset$}{
    $\operatorname{RecordAccessIfDue}(p)$\;
    \Return $p$\;
}
\tcp{Execute the following block in the bounded pool}
\If{$\operatorname{loadAcquire}(G) \ne g$}{\Return cancellation\;}
$p \leftarrow \operatorname{DecodeResizeEncode}(P,f,m,\tau)$\;
\tcp{After successful worker completion}
$\operatorname{RecordManifest}(f,m,\tau,p)$\;
$\operatorname{EnforceBudget}(B)$\;
\Return $p$\;
\end{algorithm}

\paragraph{Generation checks and their scope.}
Algorithm~\ref{alg:viewport_sched} receives the generation attached to the request rather than sampling a new generation at worker entry. Checks occur before cache processing and at the decoding worker boundary. A request that becomes stale after the second check can still finish decoding. The mechanism therefore skips obsolete pending work but does not preempt an in-progress codec operation or guarantee a particular frame rate. Cache reuse and error handling are simplified in the pseudocode.

\subsection{Thumbnail Manifest Schema and Bounded Disk Budget}
\label{sec:thumbnail_manifest}
Berry stores thumbnail indexation records inside SQLite~\cite{hipp2020sqlite} schema version \texttt{v11}:
\noindent\begin{minipage}{\columnwidth}
\begin{lstlisting}[language=SQL,basicstyle=\ttfamily\scriptsize]
CREATE TABLE thumbnail_cache_entries (
    file_id          INTEGER NOT NULL,
    modified_at      INTEGER NOT NULL,
    max_edge         INTEGER NOT NULL,
    codec            TEXT NOT NULL,
    path             TEXT NOT NULL UNIQUE,
    size_bytes       INTEGER NOT NULL,
    last_accessed_at INTEGER NOT NULL,
    PRIMARY KEY (file_id, modified_at, max_edge, codec)
) STRICT;

CREATE INDEX idx_thumbnail_cache_lru
    ON thumbnail_cache_entries(last_accessed_at, path);
\end{lstlisting}
\end{minipage}

\paragraph{Rate-Limited Access Logging:}
Updating access times on every cache hit can increase write traffic. Berry tracks recent touches in a hash map and suppresses repeated access persistence within an hour. A cached touch becomes eligible again when:
\begin{equation}
    t_{\text{current}} - t_{\text{last\_recorded}} \ge 3600 \text{ seconds}
\end{equation}
The implementation attempts to prune expired records when the map reaches 100,000 entries; this is a cleanup threshold, not a hard capacity bound. The resulting write reduction depends on the access pattern and is not measured here.

\paragraph{Tier Selection and Larger-Tier Reuse:}
Berry establishes three standard thumbnail resolution tiers: $\tau_1 = 384\text{px}$ (compact grid), $\tau_2 = 512\text{px}$ (standard masonry), and $\tau_3 = 1024\text{px}$ (detail preview). When the gallery requests tier $\tau_1$, the engine first checks whether $\tau_2$ or $\tau_3$ is already cached on disk for the same source revision ($f_{\text{id}}, m_{\text{time}}$). If present, the larger existing thumbnail is reused directly, avoiding redundant rasterization.

\subsection{Deep Pagination: Keyset Seek vs. Offset Scans}
\label{sec:keyset_pagination}
A critical scalability bottleneck in local asset indexing is deep pagination.
A basic offset query selects a page after skipping an ordered prefix:
\begin{equation}
    \mathcal{Q}_{\text{offset}} = \sigma_{\text{filter}}(\mathcal{R}) \quad \text{LIMIT } K \text{ OFFSET } N
\end{equation}
In B-Tree indexes, skipping $N$ rows requires the database engine to visit or skip $N$ preceding index entries, yielding $\mathcal{O}(N + K)$ computational complexity.

Berry addresses this by introducing sort-aware \textbf{Keyset Cursor} pagination. Let $(m_{\text{last}}, i_{\text{last}})$ represent the modification timestamp and unique file ID of the anchor record from the previous page. The subsequent page is retrieved via composite tuple comparison:
\noindent\begin{minipage}{\columnwidth}
\begin{lstlisting}[language=SQL,basicstyle=\ttfamily\scriptsize]
SELECT id, folder_id, path, container, size_bytes,
       modified_at, rating, aesthetic_score,
       is_favorite, is_nsfw, stack_id, stack_order
FROM files
WHERE folder_id = ?1
  AND (modified_at < ?2 OR (modified_at = ?2 AND id < ?3))
ORDER BY modified_at DESC, id DESC
LIMIT 400;
\end{lstlisting}
\end{minipage}

With a compatible index and query plan, cursor pagination can approach $\mathcal{O}(\log |\mathcal{R}|+K)$ work for a page of $K$ rows. The inspected migrations include \texttt{idx\_files\_cursor\_mtime} on \texttt{(modified\_at DESC, id DESC)}; this is not a folder-prefixed composite index. Folder filtering and the disjunctive cursor predicate can affect the plan, so the bound should not be assumed for every query. The benchmark uses raw SQL for this comparison and excludes cursor acquisition from the timed keyset operation. Stable ordering also does not provide snapshot consistency when records change between requests.

\subsection{Prompt-Based Burst Grouping}
\label{sec:burst_stacking}
Berry groups iterative outputs using normalized prompt fragments and a configurable time window. Fragments are split at supported delimiters, lowercased, trimmed, and deduplicated; numeric prompt-weight suffixes are removed. For nonempty fragment sets $P_A$ and $P_B$, the implementation uses ordinary Jaccard similarity~\cite{jaccard1912distribution}:
\begin{equation}
    J(P_A,P_B)=\frac{|P_A\cap P_B|}{|P_A\cup P_B|}.
\end{equation}
No inverse document frequency weighting is applied. Candidates are ordered by folder, modification time, and file ID. Each unassigned anchor collects eligible later candidates in the same folder whose similarity is at least a threshold $\theta$. When the configured time window $\Delta T$ is positive, candidates must also satisfy $t_B-t_A\le\Delta T$. Candidates with different known model names are excluded; existing stack members and empty prompts are skipped.

Grouping is greedy and anchor-based, so two members of a stack need not meet the threshold with each other. Modification time is a proxy for temporal proximity rather than a guaranteed generation timestamp. Grouping accuracy and threshold sensitivity remain to be evaluated against user judgments.

%% file: sections/05_evaluation.tex
\section{Evaluation}
\label{sec:evaluation}

We evaluate seven implementation properties: catalog ingestion and first-page access (E1), deep pagination (E2), thumbnail-manifest reconciliation and eviction (E3), virtualized gallery interaction (E4), stale-work cancellation (E5), metadata extraction (E6), and prompt grouping and vector retrieval (E7). All experiments use application commit \texttt{ae940a7a4e6} plus the archived benchmark patches. Each experiment has five independent repetitions. The reported values are medians of five run-level statistics; quantiles within a run use the nearest-rank definition. Every run package contains a machine-readable environment manifest, commands, raw standard output and error logs, sample-level CSV data, claim checks, and SHA-256 checksums.

E1 and E3 ran on an AMD Ryzen~7 7735H system with 16~GB RAM and Windows~11. E2 and E4--E7 ran on an Intel Core~i5-10400F system with 32~GB RAM and Windows~10. Both used NVMe storage and release builds. Results from different experiment families are not used for direct hardware comparisons.

\subsection{Catalog Storage and Pagination}

E1 constructs deterministic catalog records and inserts them through Berry's \texttt{upsert\_files} path in batches of 1,000. It then separates a cold database open plus first-page call from warm measurements. The API path returns at most 400 materialized records and computes the exact filtered total through \texttt{COUNT(*) OVER()}; it excludes IPC, thumbnail decoding, DOM work, and painting. Ingestion throughput was 36,656, 24,964, and 16,728 rows/s at 1,000, 10,000, and 50,000 records, respectively. Figure~\ref{fig:first_page_latency} shows that warm API latency increased with collection size because the exact window count is evaluated over the filtered set.

\input{figures/fig_first_page_latency}

E2 compares offset and keyset queries at offsets 8,000 and 40,000. The keyset anchor is the record immediately preceding the target page; every repetition asserts identical ordered IDs for the offset and keyset results. Raw-query timers materialize 400 IDs. API timers include record construction; the offset API also includes its window total. Cursor acquisition is timed separately and excluded from the speedup. As Figure~\ref{fig:pagination_comparison} shows, the 50,000-record raw keyset query was $126.4\times$ faster than raw offset traversal. API P50 latency fell from 49.52 to 2.72~ms at 10,000 records and from 254.11 to 4.90~ms at 50,000 records. Adding the 10.74~ms cursor-acquisition cost would still leave the measured 50,000-record keyset path below the raw offset time, but clients normally reuse a cursor from the preceding page.

\input{figures/fig_pagination_comparison}

\subsection{Thumbnail Cache and Gallery Interaction}

E3 creates 50,000 cache candidates under two controlled workloads: 16-byte placeholders and independently decodable WebP files. Reconciliation invokes Berry's manifest synchronizer; a second phase lowers the byte budget to 50\% of the initial footprint and invokes least-recently-used eviction. For valid WebP files, median reconciliation throughput was 15,933 files/s. Eviction removed 25,003 files in 7.83~s and brought the cache within budget in all repetitions. The placeholder control reached 13,919 files/s and removed 25,000 entries in 8.39~s. These timers measure manifest and filesystem management, rather than thumbnail encoding.

E4 runs a virtualized gallery in headless Microsoft Edge with a $1440\times900$ viewport. The deterministic page uses aspect-ratio-scaled image placeholders and records DevTools traces for cold and warm 10,000- and 50,000-item conditions. ``First usable gallery'' ends when the first visible DOM cards have layout coordinates; the scroll timer covers one visible-slice update. Table~\ref{tab:e4_gallery} reports low update costs and no measured task above 50~ms across 800 scroll steps per repetition. The accompanying object-URL audit allocated 10,000 URLs, retained the configured 3,000-entry working set, and observed 7,000 revocations in every run.

\begin{table}[t]
\centering
\caption{Controlled browser gallery measurements (E4).}
\label{tab:e4_gallery}
\resizebox{\columnwidth}{!}{%
\begin{tabular}{lrr}
\toprule
\textbf{Condition} & \textbf{First usable (ms)} & \textbf{Update P95 (ms)} \\
\midrule
10k cold & 6.5 & 0.3 \\
10k warm & 3.7 & 0.3 \\
50k cold & 9.2 & 0.3 \\
50k warm & 9.8 & 0.3 \\
\bottomrule
\end{tabular}
}
\end{table}

\subsection{Generation-Aware Cancellation}

E5 replays 20 viewport jumps over 1,600 generated 64$\times$64 PNG files. Each step requests 20 visible items and queues 60 speculative items; condition order is counterbalanced across repetitions. With generation checks enabled, all 1,200 stale speculative requests were cancelled before decode. With checks disabled, all 1,200 completed, reading 1.8~MB and consuming a median 555.45~ms of stale decode CPU time per run. Figure~\ref{fig:cancellation_ablation} shows the corresponding viewport latency. The result isolates this trace and worker configuration; it does not imply that every production workload cancels all stale work.

\input{figures/fig_cancellation_ablation}

\subsection{Metadata, Grouping, and Retrieval}

E6 invokes \texttt{berry\_metadata::extract\_metadata} on a versioned 10-case corpus: seven valid examples spanning A1111, ComfyUI, NovelAI, Fooocus, InvokeAI, and sidecar formats, plus three malformed inputs. Across 53 annotated fields, all repetitions produced 53 exact matches; the malformed cases terminated without a crash, panic, or hang. The median run-level P95 parse latency was 127.1~$\mu$s. This small curated corpus establishes regression coverage for its fixtures, rather than population-level parser accuracy.

E7 exercises Berry's implemented prompt grouping and SQLite vector search. A 40-item constructed corpus contains three prompt bursts and ten isolated items. At the default Jaccard threshold of 0.85, median F1 was 1.0 and median grouping time was 186.1~$\mu$s. The retrieval workload inserts 1,000 deterministic 512-dimensional vectors in ten deliberately separable clusters and issues 50 held-out generated queries. Recall@1 and Recall@5 were both 1.0; median P95 query latency was 2.05~ms and median insertion throughput was 1,794 vectors/s. These results verify the storage, similarity, and grouping paths under controlled ground truth. They do not measure semantic relevance on natural images or user queries.

\subsection{Threats to Validity}

The evaluation uses deterministic, purpose-built workloads on one machine per experiment family. Synthetic timestamps, prompts, images, vectors, and access traces cannot represent the full variation of personal media libraries. E4 measures a controlled gallery page with placeholders, so its timings should not be read as media decode or whole-application startup latency. E6 and E7 use small or deliberately separable ground-truth sets and therefore support regression and mechanism validation rather than broad accuracy claims. Five repetitions characterize short-run variability on the recorded systems, but do not establish performance across operating systems, storage devices, power states, or hardware classes. Finally, this work does not compare Berry with other asset managers; such a study would require matched versions, workloads, and observable timer boundaries.

%% file: figures/fig_first_page_latency.tex
% Verified E1 catalog latency across library scales.
\providecolor{berryBlue}{RGB}{43,76,146}
\providecolor{berryTeal}{RGB}{33,145,140}
\begin{figure}[t]
\centering
\begin{tikzpicture}
\begin{axis}[
    ybar=2pt,
    bar width=11pt,
    width=\columnwidth,
    height=5.0cm,
    ymin=0,
    ymax=335,
    ylabel={Latency (ms)},
    symbolic x coords={1k,10k,50k},
    xtick=data,
    enlarge x limits=0.23,
    ymajorgrids=true,
    grid style={dashed,gray!22},
    axis line style={gray!55},
    tick style={gray!55},
    axis x line*=bottom,
    axis y line*=left,
    legend style={
        at={(0.5,1.02)},
        anchor=south,
        legend columns=-1,
        font=\scriptsize,
        draw=none
    },
    legend image code/.code={\draw[#1] (0cm,-0.08cm) rectangle (0.18cm,0.08cm);},
    tick label style={font=\footnotesize},
    label style={font=\footnotesize},
    nodes near coords,
    nodes near coords style={font=\tiny,rotate=90,anchor=west},
    /pgf/number format/fixed,
    /pgf/number format/precision=1
]
\addplot[fill=berryBlue!88,draw=berryBlue] coordinates {
    (1k,8.6042) (10k,61.8444) (50k,295.7040)
};
\addplot[fill=berryTeal!78,draw=berryTeal!75!black] coordinates {
    (1k,6.5108) (10k,59.2502) (50k,285.3222)
};
\legend{Cold open + page,Warm API P50}
\end{axis}
\end{tikzpicture}
\caption{Verified E1 first-page database latency. Values are medians of five run-level statistics; the API returns at most 400 records and computes an exact filtered total.}
\label{fig:first_page_latency}
\end{figure}
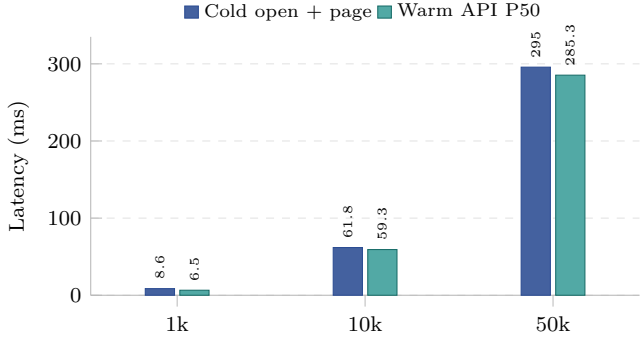

%% file: figures/fig_pagination_comparison.tex
% Verified E2 raw-query pagination comparison.
\providecolor{berryBlue}{RGB}{43,76,146}
\providecolor{berryTeal}{RGB}{33,145,140}
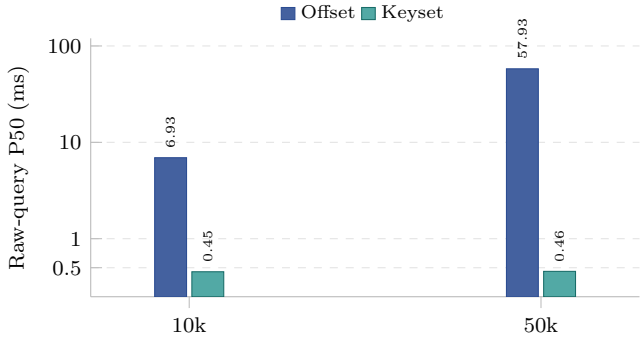
\begin{figure}[t]
\centering
\begin{tikzpicture}
\begin{axis}[
    ybar=2pt,
    bar width=12pt,
    width=\columnwidth,
    height=5.0cm,
    ymode=log,
    log origin=infty,
    ymin=0.25,
    ymax=120,
    ytick={0.5,1,10,100},
    yticklabels={0.5,1,10,100},
    ylabel={Raw-query P50 (ms)},
    symbolic x coords={10k,50k},
    xtick=data,
    enlarge x limits=0.28,
    ymajorgrids=true,
    grid style={dashed,gray!22},
    axis line style={gray!55},
    tick style={gray!55},
    axis x line*=bottom,
    axis y line*=left,
    legend style={
        at={(0.5,1.02)},
        anchor=south,
        legend columns=-1,
        font=\scriptsize,
        draw=none
    },
    legend image code/.code={\draw[#1] (0cm,-0.08cm) rectangle (0.18cm,0.08cm);},
    tick label style={font=\footnotesize},
    label style={font=\footnotesize},
    nodes near coords,
    nodes near coords style={font=\tiny,rotate=90,anchor=west},
    point meta=explicit symbolic
]
\addplot[fill=berryBlue!88,draw=berryBlue] coordinates {
    (10k,6.9266) [6.93] (50k,57.9284) [57.93]
};
\addplot[fill=berryTeal!78,draw=berryTeal!75!black] coordinates {
    (10k,0.4540) [0.45] (50k,0.4583) [0.46]
};
\legend{Offset,Keyset}
\end{axis}
\end{tikzpicture}
\caption{Verified E2 deep-page raw-query latency for 400 IDs. Keyset pagination was $15.3\times$ faster at 10,000 records and $126.4\times$ faster at 50,000 records. Cursor acquisition is measured separately and excluded.}
\label{fig:pagination_comparison}
\end{figure}

%% file: figures/fig_cancellation_ablation.tex
% Verified E5 cancellation ablation.
\providecolor{berryBlue}{RGB}{43,76,146}
\providecolor{berryTeal}{RGB}{33,145,140}
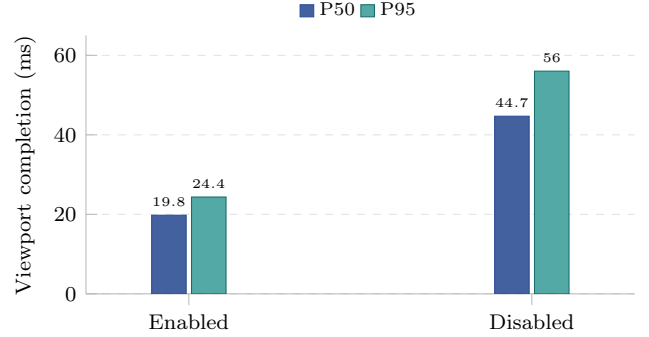
\begin{figure}[t]
\centering
\begin{tikzpicture}
\begin{axis}[
    ybar=2pt,
    bar width=13pt,
    width=\columnwidth,
    height=5.0cm,
    ymin=0,
    ymax=65,
    ylabel={Viewport completion (ms)},
    symbolic x coords={Enabled,Disabled},
    xtick=data,
    enlarge x limits=0.3,
    ymajorgrids=true,
    grid style={dashed,gray!22},
    axis line style={gray!55},
    tick style={gray!55},
    axis x line*=bottom,
    axis y line*=left,
    legend style={
        at={(0.5,1.02)},
        anchor=south,
        legend columns=-1,
        font=\scriptsize,
        draw=none
    },
    legend image code/.code={\draw[#1] (0cm,-0.08cm) rectangle (0.18cm,0.08cm);},
    tick label style={font=\footnotesize},
    label style={font=\footnotesize},
    nodes near coords,
    nodes near coords style={font=\tiny},
    /pgf/number format/fixed,
    /pgf/number format/precision=1
]
\addplot[fill=berryBlue!88,draw=berryBlue] coordinates {
    (Enabled,19.7780) (Disabled,44.6816)
};
\addplot[fill=berryTeal!78,draw=berryTeal!75!black] coordinates {
    (Enabled,24.3604) (Disabled,56.0223)
};
\legend{P50,P95}
\end{axis}
\end{tikzpicture}
\caption{Verified E5 cancellation ablation. Generation checks prevented all 1,200 stale speculative decodes in the tested trace; without checks, all 1,200 completed. Values are medians of five run-level quantiles.}
\label{fig:cancellation_ablation}
\end{figure}

%% file: sections/06_open_source.tex
\section{Artifact Availability and Reproducibility}
\label{sec:open_source}

Berry AI Studio is released under the GNU Affero General Public License version~3 at:
\begin{center}
\url{https://github.com/BerryUIKI/Omera}
\end{center}
The experiments use application commit \href{https://github.com/BerryUIKI/Omera/commit/ae940a7a4e6d9249ff2f201ba0f6b77a1256e87b}{\texttt{ae940a7a4e6d}}. Benchmark-only source patches are stored with each experiment so that the measured code path can be reconstructed without treating instrumentation as application functionality.

The paper repository contains the E1--E7 runners and 35 immutable run directories under \texttt{experiments/runs}. Each directory records the source revision, environment, workload, timer boundary, condition order, and quantile method in \texttt{manifest.json}; it also includes the exact commands, raw logs, sample-level results, claim assertions, supporting artifacts, and a checksum inventory. The packages can be checked and aggregated with:
\noindent\begin{minipage}{\columnwidth}
\begin{lstlisting}[language=bash,basicstyle=\ttfamily\scriptsize]
python experiments/validate_run.py RUN_DIRECTORY
python experiments/analyze_release_results.py
\end{lstlisting}
\end{minipage}

The second command regenerates \texttt{release\_results.json}, which is the numerical source for the tables in Section~\ref{sec:evaluation}. Browser trace archives from E4 are retained alongside the derived frame samples. This separation between raw observations, aggregation, and prose allows reported values to be audited and recalculated without parsing the manuscript.

%% file: sections/07_conclusion.tex
\section{Conclusion}
\label{sec:conclusion}

Berry AI Studio combines heterogeneous provenance extraction, local catalog storage, virtualized browsing, bounded thumbnail management, and on-device retrieval in a desktop system. Its central design choice is to keep expensive work aligned with the current interaction: cursor queries avoid repeated deep traversal, viewport generations invalidate obsolete thumbnail requests, and a persistent manifest makes cache state explicit.

The evaluation tests these mechanisms on a pinned implementation with five independent repetitions. At 50,000 records, keyset pagination reduced median raw deep-page latency from 57.93 to 0.458~ms, while the cancellation ablation prevented 1,200 stale decodes and reduced median viewport completion from 44.68 to 19.78~ms in the controlled trace. Checksummed artifacts also cover browser virtualization, cache reconciliation, metadata fixtures, prompt grouping, and vector search. The reported workloads are intentionally bounded; broader claims about semantic relevance, heterogeneous devices, and naturally occurring libraries require larger external corpora and multi-system evaluation. The released implementation and evidence packages provide a concrete basis for that work.

%% file: main.bbl
\begin{thebibliography}{10}

\bibitem{rombach2022high}
R.~Rombach, A.~Blattmann, D.~Lorenz, P.~Esser, and B.~Ommer, ``High-resolution
  image synthesis with latent diffusion models,'' in {\em Proceedings of the
  IEEE/CVF Conference on Computer Vision and Pattern Recognition (CVPR)},
  pp.~10684--10695, 2022.

\bibitem{comfyui2023}
Comfyanonymous, ``{ComfyUI}: A powerful and modular stable diffusion {GUI} and
  backend with a graph/nodes interface.''
  \url{https://github.com/comfyanonymous/ComfyUI}, 2023.

\bibitem{sdwebui2022}
{AUTOMATIC1111}, ``{Stable Diffusion web UI}: Modular browser interface for
  {Stable Diffusion}.''
  \url{https://github.com/AUTOMATIC1111/stable-diffusion-webui}, 2022.

\bibitem{fooocus2023}
{lllyasviel}, ``{Fooocus}: Focus on prompting and generating.''
  \url{https://github.com/lllyasviel/Fooocus}, 2023.

\bibitem{digikam2010}
{digiKam Team}, ``{digiKam}: Photo management software.''
  \url{https://www.digikam.org}.
\newblock Software project.

\bibitem{eagle2020}
{Eagle App}, ``{Eagle}: Organize all your reference images in one place.''
  \url{https://eagle.cool}, 2020.

\bibitem{rayon2023}
N.~Matsakis, J.~Stone, {\em et~al.}, ``Rayon: A data-parallelism library for
  {Rust}.'' \url{https://github.com/rayon-rs/rayon}, 2023.

\bibitem{radford2021learning}
A.~Radford, J.~W. Kim, C.~Hallacy, A.~Ramesh, G.~Goh, S.~Agarwal, G.~Sastry,
  A.~Askell, P.~Mishkin, J.~Clark, {\em et~al.}, ``Learning transferable visual
  models from natural language supervision,'' in {\em International Conference
  on Machine Learning (ICML)}, pp.~8748--8763, PMLR, 2021.

\bibitem{zhai2023sigmoid}
X.~Zhai, B.~Mustafa, A.~Kolesnikov, and L.~Beyer, ``Sigmoid loss for language
  image pre-training,'' in {\em Proceedings of the IEEE/CVF International
  Conference on Computer Vision (ICCV)}, pp.~11975--11986, 2023.

\bibitem{onnxruntime2021}
{Microsoft Corporation}, ``{ONNX Runtime}: cross-platform, high performance
  scoring engine for {ML} models.'' \url{https://onnxruntime.ai}, 2021.

\bibitem{tauri2024}
{Tauri Programme}, ``{Tauri 2.0}: Build smaller, faster, and more secure
  desktop and mobile applications with a web frontend.''
  \url{https://tauri.app}, 2024.

\bibitem{matsakis2014rust}
N.~D. Matsakis and F.~S. Klock~II, ``The {Rust} language,'' {\em ACM SIGAda
  Ada Letters}, vol.~34, no.~3, pp.~103--104, 2014.

\bibitem{hipp2020sqlite}
{SQLite Project}, ``{SQLite}: About sqlite.''
  \url{https://www.sqlite.org/about.html}.
\newblock Accessed September 21, 2026.

\bibitem{jaccard1912distribution}
P.~Jaccard, ``The distribution of the flora in the alpine zone,'' {\em New
  Phytologist}, vol.~11, no.~2, pp.~37--50, 1912.

\end{thebibliography}
